\documentstyle{article}
\title{\vspace*{-1.cm}
\hspace*{9.6cm} {\large gr-qc/9506038\vspace*{1.7cm}}\\
\Large{{\bf Involutions on the Algebra of Physical Observables from
Reality Conditions\vspace*{.9cm}}}}
\author{
Guillermo A. Mena Marug\'an\vspace*{.5cm}\\
Instituto de Matem\'aticas y F\'{\i}sica
Fundamental, C.S.I.C.,\\ Serrano 121, 28006 Madrid, Spain.}
\date{}

\begin{document}

\maketitle
\large
\setlength{\baselineskip}{.825cm}
\vspace*{.55cm}

\begin{center}
{\bf Abstract}
\end{center}
\vspace*{.4cm}

Some aspects of the algebraic quantization programme proposed by Ashtekar
are revisited in this article. It is proved that, for systems with
first-class constraints, the involution introduced on the algebra of quantum
operators via reality conditions can never be projected unambiguously to the
algebra of physical observables, ie, of quantum observables modulo
constraints. It is nevertheless shown that, under sufficiently general
assumptions, one can still induce an involution on the algebra of
physical observables from reality conditions, though the involution
obtained depends on the choice of particular representatives for
the equivalence classes of quantum observables and this implies an
additional ambiguity in the quantization procedure suggested by Ashtekar.

\vspace*{1.cm}


\newpage
\renewcommand{\thesection}{\Roman{section}.}
\renewcommand{\theequation}{\arabic{section}.\arabic{equation}}
\renewcommand{\thefootnote}{a}

\section {Introduction}

Recently, Ashtekar {\it et al}$^{\;1-3}$ have ellaborated a programme for the
non-perturbative quantization of dynamical systems with first-class
constraints. This programme is specially designed to deal with the problem
of quantizing general relativity, and has already been carried out
successfully in a number of lower dimensional gravita\-tional models,
including minisuperspaces,$^{4,5}$ midisuperspaces$^6$ and 2+1
gravity.$^{1,3,7}$ The programme proposed by Ashtekar is an
extension, based on the algebraic ap\-proach to quantum mechanics,$^8$
of Dirac's canonical quantization method.$^9$ One of the main novelties with
respect to Dirac's procedure is the introduction of a
prescription to find the inner product in the space of quantum states.
This allows one to adhere to the standard probabilistic interpretation of
quantum mechanics when the quantization can be achieved.

Ashtekar's programme
consists of a series of steps that, after completion, should
provide us with a consistent quantum theory. It can be applied, in principle,
to any classical system whose phase space $\Gamma$ is a real symplectic
manifold.$^1$

One must first choose a subspace $S$ of the vector space of smooth
complex functions on $\Gamma$. This subspace must contain the unit function
and be closed both under complex conjugation and Poisson brackets.$^2$
In addition, $S$ has to be complete, in the sense that any sufficiently
regular complex function on phase space should be expressable as a sum of
products of elements in $S$ (or as a limit of this type of sums).$^2$

Each element $X$ in $S$ is to be regarded as an elementary classical variable
which is unambiguously associated with an abstract operator $\hat{X}$.
One then constructs
the free associative algebra generated by these elementary quantum
operators. On this algebra, one imposes the commutation relations that follow
from the classical Poisson brackets, namely, if $X,Y\in S$, one must demand
that $[\hat{X},\hat{Y}]=i\hbar \widehat{\{X,Y\}}$ (at least up to terms
proportional to $\hbar^2$). If there exist algebraic relations between the
elements in $S$ (eg, when the dimension of $S$ is greater than that of
$\Gamma$), such relations have also to be imposed on the corresponding
quantum operators, with a suitable choice of factor ordering, if needed.$^2$
The algebra of operators obtained in this way will be called
${\cal{A}}$.

At this point one should promote the complex conjugation
relations in $S$ to an involution on ${\cal{A}}$. We recall that an involution
$\star$ on the algebra ${\cal A}$ is a map $\star:\;{\cal A}\rightarrow
{\cal A}$ that satisfies
\begin{equation} (\hat{X}^{\star})^{\star}=\hat{X}\;,  \end{equation}
\begin{equation} (\hat{X}+\lambda \hat{Y})^{\star}= \hat{X}^{\star}+
\overline{\lambda}\;\hat{Y}^{\star}\;,\;\;\;\;\;\; (\hat{X}\hat{Y})^{\star}=
\hat{Y}^{\star}\hat{X}^{\star}\;,\end{equation}
for all $\hat{X},\hat{Y}\in {\cal A}$ and complex numbers $\lambda$. Here,
$\overline{\lambda}$ is the complex conjugate to $\lambda$. To introduce the
desired involution on ${\cal A}$,
one can proceed in the following manner. For every $X,Y\in S$ such that
$Y$ is the complex conjugate to $X$, define $\hat{X}^{\star}=\hat{Y}$, and
use properties (1.2) to extend this definition to all the operators
in ${\cal A}$. It is not difficult to check then that one gets an
involution on ${\cal A}$ provided that the $\star$-operation is compatible
with the structure of this algebra. This amounts to require that the
commutation and algebraic relations between elementary operators which have
been imposed on ${\cal A}$ are stable under the $\star$-operation, in the
sense that their $\star$-conjugates do not supply any new relation
which is not implied by the original ones. We will assume hereafter
that this is in fact the case, and denote the resulting $\star$-algebra by
${\cal A}^{(\star)}$. The $\star$-relations in ${\cal A}^{(\star)}$ are
usually called reality conditions,$^1$ for they capture the complex
conjugation relations between elementary classical variables.

The next step in the quantization consists in finding a faithful
representation for the abstract algebra ${\cal A}$ by linear operators
acting on a complex vector space $V$. If the classical system possesses
first-class constraints $\{C_i\}$, these constraints must now be explicitly
represented by operators $\{\hat{C}_i\}$. In general, a choice of factor
ordering, and of regularization in infinite dimensional systems,$^{2,3}$
are needed at this point in order to get a consistent algebra of quantum
constraints,$^9$ that is, to guarantee that
\begin{equation} [\hat{C}_i,\hat{C}_j]=\hat{f}_{ij}^{\;\;\,k}\hat{C}_k\;,
\end{equation}
where $\hat{f}_{ij}^{\;\;\;k}\in {\cal A}$ and we use the convention that
pairs of contracted indices are summed over.

The kernel $V_{p}\subset V$ of the constraints $\{\hat{C}_i\}$ supplies
the vector subspace of quantum states. One must then determine the
subalgebra ${\cal A}_p\subset {\cal A}$  of operators which leave $V_p$
invariant. These operators commute weakly with the quantum constraints,
\begin{equation} \hat{A}\in {\cal A}_p\;\;\;\Longleftrightarrow\;\;\;
[\hat{A},\hat{C}_i]=\hat{h}_i^{\;\,j}\hat{C}_j\;\;\;\;\;\;(\hat{h}_i^{\,\;j}
\in {\cal A})\;.\end{equation}

Let us define now
\begin{equation} {\cal I}_C\equiv \{ \hat{X}^i\hat{C}_i;\;\hat{X}^i\in{\cal A}
\}\;.\end{equation}
Using Eqs. (1.3,4) one can show that ${\cal I}_C\subset{\cal A}_p$ and
that, $\forall \hat{I}\in {\cal I}_C$ and $\forall \hat{A}\in {\cal A}_p$,
both $\hat{A}\hat{I}$ and $\hat{I}\hat{A}$
belong to ${\cal I}_C$, so that ${\cal I}_C$ is an ideal of
${\cal A}_p$. On the other hand, if $\hat{A}\in {\cal A}_p$, all the operators
of the form $\hat{B}=\hat{A}+\hat{I}$, with $\hat{I}\in {\cal I}_C$, have
exactly the same action on quantum states, for $V_p$ is anihilated by the
quantum constraints. In order to obtain the algebra ${\cal A}^{\prime}_p$
of operators with a well-defined action on $V_p$, one should therefore
take the quotient of ${\cal A}_p$ by the ideal ${\cal I}_C\,$:
\begin{equation} {\cal A}^{\prime}_p\equiv {\cal A}_p/{\cal I}_C\;.
\end{equation}
The operators in ${\cal A}^{\prime}_p$ are the quantum physical observables
of the system.$^2$

The quantization programme presented so far leaves
a certain freedom in the following steps: a) the selection of the subspace
$S$ of elementary classical variables,\linebreak
b) the construction of the linear
representation for the algebra ${\cal A}$ of quantum oper\-ators, and c) the
choice of factor ordering in the quantum constraints $\{\hat{C}_i\}$. The
final result of the quantization process will depend on these inputs.$^2$
In particular, Ashtekar and Tate$^2$ assumed at this stage that, with a
judicious choice of such inputs (and at least for a large variety of physical
systems), the involution defined on
${\cal A}^{(\star)}$ would unambiguously induce an involution on
${\cal A}^{\prime}_p$. Actually, the $\star$-relations
will project unambiguously to the algebra of
physical observables only if two conditions are fulfilled.
On the one hand, ${\cal A}_p\subset {\cal A}$ must be invariant
under the $\star$-operation: $\forall \hat{A}\in {\cal A}_p$,
$\hat{A}^{\star}\in {\cal A}_p$. On the other hand, it is necessary that
${\cal I}_C\subset {\cal A}_p$ be a $\star$-ideal of ${\cal A}_p$:
$\forall \hat{I}\in{\cal I}_C$, $\hat{I}^{\star}\in {\cal I}_C$.
When this is the case, the $\star$-operation provides a uniquely defined map
between equivalence classes in ${\cal A}^{\prime}_p$ which satisfies the
properties (1.1,2) of an involution. Such an
involution will be denoted again by $\star$, and the resulting
$\star$-algebra of physical observables by ${\cal A}^{\prime\;\,(\star)}_p$.

The idea suggested by Ashtekar$^{1-3}$ is to employ the involution on
${\cal A}^{\prime\;\,(\star)}_p$ to select the inner product $<,>$
on $V_p$ and, therefore, the Hilbert space ${\cal H}$ of physical states
(normalizable quantum states). More specifically, he proposed to determine
the inner product on $V_p$ by demanding that the $\star$-relations between
physical observables are realized as adjoint relations on the Hilbert space
${\cal H}$, ie,
\begin{equation} <\Psi,\hat{A}^{\prime}\Phi>=<\hat{B}^{\prime}\Psi,\Phi>\;\;\;
\;\;\forall\,\Phi,\Psi\!\in\!{\cal H},\;\;\;
\forall\hat{A}^{\prime},\hat{B}^{\prime}\!=\!(\hat{A}^{\prime})^{\star}\!\in
{\cal A}^{\prime\;\,(\star)}_p.\end{equation}
Rendall showed$^{10}$ that this condition is such a severe restriction on the
inner product that, if an admissible inner product exists, it is unique
(up to a positive global factor) under very general assumptions.

This completes the quantization programme put forward by Ashtekar.
If this programme can be carried out for a given classical system, one would
arrive at a mathematically consistent quantum theory in which real physical
observables would be represented by self-adjoint operators acting on a
Hilbert space of physical states.

The purpose of this work is to demonstrate however that one of the
steps of the above quantization method can never be achieved. We will
prove in Section II that the $\star$-relations in ${\cal A}^{(\star)}$
can never be projected unambiguously
to the algebra of phys\-ical observables. This problem can
be nonetheless overcome by slightly modifying Ashtekar's programme, as we
will show in Section III. The price to be paid is to allow a new
freedom in the quantization process. A particular procedure to introduce
an involution on ${\cal A}^{\prime}_p$ from reality conditions should then
be adopted. The subtleties
that arise in defining such an involution are illustrated in Section IV
by considering some simple physical systems. We finally
discuss the physical implications of our results and conclude
in Section V.

\section{Ambiguities in the Reality Conditions on Physical\newline
Observables}
\setcounter{equation}{0}

We want to prove that reality conditions (ie, the $\star$-relations between
quantum operators) never project unambiguously to
the algebra of physical observables when there exist first-class
constraints on the system. We will assume that the faithful, linear
representation constructed for the algebra ${\cal A}$ of quantum operators is
irreducible. Otherwise, one should decompose it in irreducible
components, and apply the proof to follow to each component
separately.

We have seen that, in order to obtain a uniquely defined involution on
physical observables from reality conditions, it is necessary that both
${\cal A}_p$ and ${\cal I}_C$ be invariant under the $\star$-operation.
In particular, we should have
\begin{equation} \forall \hat{I}\in{\cal I}_C,\;\;\;
\hat{I}^{_0}\equiv\hat{I}^{\star}\in{\cal I}_C\;.\end{equation}
Taking $\hat{I}$ equal to each of the quantum constraints
and recalling definiton (1.5), we hence get
\begin{equation} \hat{C}^{\star}_i=\hat{Y}_i^{\,j}\hat{C}_j\;,\end{equation}
for some $\hat{Y}_i^{\,j}\in {\cal A}$.
Select now one of the quantum constraints, eg, $\hat{C}_1$, and consider
all the operators of the form $\hat{I}_1=\hat{Z}\hat{C}_1\in{\cal I}_C$,
with $\hat{Z}\in{\cal A}$. Employing again condition (2.1), and using Eq.
(2.2), we obtain
\begin{equation} (\hat{I}_1)^{\star}=\hat{C}_1^{\star}\hat{Z}^{\star}=
\hat{Y}_1^{\,j}\hat{C}_j\hat{Z}^{\star}\equiv
\hat{I}_1^{_0}=\hat{X}_1^{\,k}\hat{C}_k\;,\end{equation}
where we have expressed $\hat{I}_1^{_0}\in {\cal I}_C$
as a combination of quantum constraints.

On the other hand, the image $\hat{Z}^{\star}$ of
all the operators $\hat{Z}\in {\cal A}$ is again the whole algebra
${\cal A}$, because the $\star$-operation is an involution. Relation
(2.3) therefore implies that, $\forall \hat{Z}\in {\cal A}$, there exist
$\hat{X}_1^{\,k}\in {\cal A}$ such that
\begin{equation} \hat{Y}_1^{\,j}\hat{C}_j\hat{Z}=
\hat{X}_1^{\,k}\hat{C}_k\;.\end{equation}
This identity between operators must hold on any element of $V$, the vector
space on which ${\cal A}$ has been represented.
Choosing then $\Phi\in V_p\subset V$ with $\Phi$ different from
zero, it follows from Eq. (2.4) that, $\forall \hat{Z}\in {\cal A}$,
\begin{equation} \hat{Y}_1^{\,j}\hat{C}_j(\hat{Z}\Phi)=\hat{X}_1^{\,k}
\hat{C}_k\Phi=0\;,\end{equation}
for the physical state $\Phi$ is anihilated by all quantum
constraints. Besides, since the representation constructed is irreducible
and $\Phi\neq 0$, the range of $\hat{Z}\Phi\,$
($\forall \hat{Z}\in {\cal A}$) must be the whole vector space $V$.
So, the above equation states that $V$ is the kernel of the operator
$\hat{Y}_1^{\,j}\hat{C}_j$. Being the representation for
${\cal A}$ faithful, we then must have
\begin{equation} \hat{Y}_1^{\,j}\hat{C}_j=\hat{0}\;.\end{equation}
But this is clearly inconsistent with the fact that the $\star$-operation
is an involution, because, using Eqs. (2.2) and (2.6), we get that
$\hat{C}_1=\!(\hat{C}_1^{\star})^{\star}\!=\hat{0}$.
In this way, we conclude that, when there exist first-class constraints,
${\cal I}_C$ is never a $\star$-ideal of ${\cal A}_p$ and, there\-fore,
reality conditions do not project unambiguously
to the algebra of physical observables.

Thus, the $\star$-operation never provides a uniquely defined
map between equivalence classes in ${\cal A}^{\prime}_p$.
Moreover, even though one could find a representative $\hat{A}$ for a given
physical observable $\hat{A}^{\prime}$
such that $\hat{A}^{\star}\in {\cal A}_p$, it is not yet true that the
$\star$-conjugates of all the operators in the equivalence class
$\hat{A}^{\prime}$ (ie, the operators $\hat{A}+\hat{I}$, with $\hat{I}\in
{\cal I}_C$) belong at least to the algebra ${\cal A}_p$.

For the sake of an example, let us consider a classical system whose
phase space admits a set of global coordinates of the form
$s\equiv\{t,H,x,p\}$, with $t,H,x,p\in I\!\!\!\,R$, and $H$ and $p$
the momenta canonically conjugate to $t$ and $x$, respectively.
Suppose, in addition, that there exists only one first-class constraint
on the system, given by $H=0$. This extremely simple example describes, for
instance, a Kantowski-Sachs model with positive cosmological constant.$^5$

As elementary classical variables, we can choose the complex vector
space spanned by $s$ and the unity. The $\star$-operation on the
corresponding algebra ${\cal A}$ of quantum operators is defined by
\begin{equation}\hat{t}^{\star}=\hat{t}\;,\;\;\;\;\;\;\;\;\hat{H}^{\star}
=\hat{H}\;,\end{equation}
\begin{equation} \hat{x}^{\star}=\hat{x}\;,\;\;\;\;\;\;\hat{p}^{\star}
=\hat{p}\;,\;\;\;\;\;\;\hat{1}^{\star}=\hat{1}\;,\end{equation}
and the properties (1.2) of an involution. The only quantum constraint
is $\hat{H}=0$. On the other hand, it is not difficult to prove that
the equivalence classes in ${\cal A}^{\prime}_p$ of the operators $\hat{1}$,
$\hat{x}$ and $\hat{p}$ form a complete set of physical observables.
Using Eq. (2.8), it then follows that each equivalence
class of observables possesses at least a representative whose
$\star$-conjugate belongs to the algebra ${\cal A}_p$.
However, the $\star$-image of different representatives do not coincide
in general (not even modulo the constraint $\hat{H}=0$).
Let us take, for instance, the operators $\hat{x}$, $\hat{x}+\hat{t}\hat{H}$
and $\hat{x}+(\hat{t})^2\hat{H}$, all of them in the same equivalence class
of physical observables. From Eqs. (2.7,8) and the commutator
$[\hat{t},\hat{H}]=i\hbar\hat{1}$, we get
\begin{equation} \hat{x}^{\star}\!=\hat{x},\;\;\;\left(\hat{x}+\hat{t}\hat{H}
\right)^{\star}\!\!=\!\left(\hat{x}+\hat{t}\hat{H}\right)\!-i\hbar\hat{1},\;
\;\;\left(\hat{x}+(\hat{t})^2\hat{H}\right)^{\star}\!\!=\!\left(\hat{x}+
(\hat{t})^2\hat{H}\right)\!-2i\hbar\hat{t}.\end{equation}
Hence, the $\star$-conjugate to $\hat{x}$ and to $\hat{x}+\hat{t}\hat{H}$
belong to different classes of observables, whereas the $\star$-conjugate
to $\hat{x}+(\hat{t})^2\hat{H}$ is not even in ${\cal A}_p$.

\section{Involutions on Physical Observables}
\setcounter{equation}{0}

We have seen that the $\star$-relations in ${\cal A}^{(\star)}$ do not
project unambiguously to ${\cal A}^{\prime}_p$, because the
$\star$-operation never maps all the representatives of a class of physical
observables into another equivalence class. In order to define the
$\star$-conjugate to a physical observable, one is therefore forced to
choose first a particular representative for it. We now want to discuss
under which circumstances it is possible to introduce an involution
on ${\cal A}^{\prime}_p$ by this procedure, namely, by selecting a
particular representative for each equivalence class in ${\cal A}^{\prime}_p$.

To construct an involution $\star$ on ${\cal A}^{\prime}_p$, it
actually suffices to define the $\star$-operation on an (over-)complete
set of physical observables, and demand that this operation verifies
conditions (1.2). Suppose then that $\{\hat{U}^{\prime}_a\}$ is a
complete set in ${\cal A}^{\prime}_p$, that is, that ${\cal A}^{\prime}_p$
can be obtained from the free associative algebra ${\cal B}^{\prime}$
generated by $\{\hat{U}^{\prime}_a\}$ by imposing the commutation
relations between the observables $\hat{U}^{\prime}_a$, as well as any
algebraic relation that could exist between them. Assume also that one
can find rep\-resentatives $\{\hat{U}_a\}$ of the observables
$\{\hat{U}^{\prime}_a\}$ such that their $\star$-conjugates
$\{\hat{U}^{\star}_a\}$ belong to ${\cal A}_p$. One might then hope that the
$\star$-operation on ${\cal A}^{\prime}_p$ could be defined by
\begin{equation} (\hat{U}^{\prime}_a)^{\star}=(\hat{U}^{\star}_a)^{\prime}\;,
\end{equation}
where $(\hat{U}^{\star}_a)^{\prime}$ denotes the equivalence class of
$\hat{U}^{\star}_a$. However, we will prove that the assumptions introduced
above do not guarantee that Eq. (3.1) leads to a well-defined involution
on the algebra of physical observables.

The proof makes use of the fact that, being $\{\hat{U}^{\prime}_a\}$ complete
in ${\cal A}^{\prime}_p$, any operator in the algebra ${\cal A}_p$ should be
expressable, modulo an element in the ideal ${\cal I}_C$ (1.5), as (possibly
a limit of) a sum of products of the representatives $\{\hat{U}_a\}$.
In particular, since every $\hat{U}^{\star}_a\in {\cal A}_p$, one gets
\begin{equation} \hat{U}^{\star}_a=\sum_n \lambda_a^{\,b_1...b_n}
\hat{U}_{b_1}...\hat{U}_{b_n}+\hat{X}_a^{\,i}\hat{C}_i\;,\end{equation}
with $\hat{X}_a^{\,i}\in {\cal A}$ and the $\lambda_a^{\,b_1...b_n}$'s
some complex numbers. Hence, from Eq. (3.1),
\begin{equation} (\hat{U}^{\prime}_a)^{\star}=\sum_n \lambda_a^{\,b_1...b_n}
\hat{U}^{\prime}_{b_1}...\hat{U}^{\prime}_{b_n}.\end{equation}
This $\star$-operation will be an involution on ${\cal A}^{\prime}_p$
only if $((\hat{U}^{\prime}_a)^{\star})^{\star}=\hat{U}^{\prime}_a$
for all $\hat{U}^{\prime}_a$. This, together
with Eqs. (1.2), (3.1) and (3.3), implies
\begin{equation} \hat{U}^{\prime}_a=\sum_n \overline{\lambda}_a
^{\,b_1...b_n} (\hat{U}^{\star}_{b_n})^{\prime}...
(\hat{U}^{\star}_{b_1})^{\prime}.\end{equation}
On the other hand, we have from Eq. (3.2)
\begin{equation} \hat{U}_a=\sum_n \overline{\lambda}_a^{\,b_1...b_n}
\hat{U}^{\star}_{b_n}...\hat{U}^{\star}_{b_1}+\hat{C}^{\star}_i
\hat{X}_a^{\,i\,\star},\end{equation}
since the $\star$-operation is an involution on ${\cal A}^{(\star)}$.
Consistency of Eq. (3.4) with (3.5) requires then
\begin{equation} \hat{C}^{\star}_i\hat{X}_a^{\,i\,\star}=\hat{Y}_a^{\,i}
\hat{C}_i,\end{equation}
for some operators $\hat{Y}_a^{\,i}\in {\cal A}$. This condition will not be
satisfied by generic operators $\hat{X}_a^{\,i}\hat{C}_i\in {\cal I}_C$,
because the ideal ${\cal I}_C$ is not invariant under the
$\star$-operation when there exist first-class constraints on the system.
Therefore, the $\star$-relations (3.3) will not supply
in general an involution on ${\cal A}^{\prime}_p$.
To obtain that involution, it is necessary that both conditions
(3.2) and (3.6) are satisfied by the representatives of our
complete set of physical observables.

We will study now the case in which these requirements hold
for our particular choice of representatives. Our previous
discussion shows that the $\star$-operation defined by
Eqs. (3.3) and (1.2) is then an involution on ${\cal B}^{\prime}$, the free
associative algebra generated by $\{\hat{U}^{\prime}_a\}$. Recalling that
the algebra ${\cal A}^{\prime}_p$ of physical observables can be obtained
from ${\cal B}^{\prime}$ by imposing on its generators the commutation
relations and any existing algebraic relations, we conclude that the
$\star$-operation introduced on ${\cal B}^{\prime}$ straightforwardly supplies
an involution on ${\cal A}^{\prime}_p$ provided that such an operation is
compatible with the relations imposed on the generators
$\{\hat{U}^{\prime}_a\}$.
In other words, the $\star$-conjugate to those relations should not lead
to any new restriction on ${\cal B}^{\prime}$. When this requisite
is fulfilled, one gets an involution on ${\cal A}^{\prime}_p$
which captures the reality conditions on quantum operators.

Notice that the involution at which one arrives depends, nevertheless, on two
choices: the complete set of physical observables and the representatives
for them. In general, distinct choices may lead to different
involutions on the algebra of physical observables. We will comment on this
point further in Section V.

A situation which is often encountered in physical applications$^{4,5}$
is that one can find a complete set in ${\cal A}^{\prime}_p$ admitting
representatives $\{\hat{U}_a\}$ such that the complex vector space
spanned by them is closed under reality conditions, ie,
\begin{equation} \hat{U}^{\star}_a=\lambda_a^{\,b}\hat{U}_b\;.\end{equation}
In this case, assumption (3.2) holds with $\hat{X}_a^{\,i}\hat{C}_i=\hat{0}$,
so that Eq. (3.6) is trivially satisfied. It is then at least possible to
obtain an involution on the free algebra ${\cal B}^{\prime}$
by replacing the operators $\hat{U}_a$ in Eq. (3.7)
with their corresponding equivalence classes of physical observables.

\section{Examples}
\setcounter{equation}{0}

Let us illustrate our discussion by dealing with some examples.
Consider, for in\-stance, the physical system that was analysed at the
end of Section II. A complete set of physical observables for this
system is ${\cal O}^{\prime}\equiv \{\hat{1}^{\prime},\hat{x}^{\prime},
\hat{p}^{\prime}\}$, where $\hat{1}^{\prime}$, $\hat{x}^{\prime}$ and
$\hat{p}^{\prime}$ are the equivalence classes of the operators $\hat{1}$,
$\hat{x}$ and $\hat{p}$, respectively. We can select these operators as the
representatives of ${\cal O}^{\prime}$. The associated reality conditions,
which are given by Eq. (2.8), have the form (3.7). So, hypotheses (3.2) and
(3.6) apply. We can therefore try to induce an involution on
${\cal A}^{\prime}_p$ by the procedure explained in Section III. Since there
exist no algebraic relations in ${\cal O}^{\prime}$, the only consistency
requirement that must be satisfied in order to get the desired involution
is that reality conditions (2.8) are compatible with the commutators
of the physical observables in ${\cal O}^{\prime}$. There is just one
commutator different from zero: $[\hat{x}^{\prime},\hat{p}^{\prime}]=
i\hbar\hat{1}^{\prime}$. On the other hand, we obtain from Eqs. (2.8)
and (3.3)
\begin{equation} (\hat{x}^{\prime})^{\star}=\hat{x}^{\prime}\;,\end{equation}
\begin{equation} (\hat{p}^{\prime})^{\star}=\hat{p}^{\prime}\,,\;\;\;\;\;\;\;
(\hat{1}^{\prime})^{\star}=\hat{1}^{\prime}\;.\end{equation}
Taking then the $\star$-conjugate to $[\hat{x}^{\prime},\hat{p}^{\prime}]$,
we get
\begin{equation} \left([\hat{x}^{\prime},\hat{p}^{\prime}]\right)^{\star}=
[(\hat{p}^{\prime})^{\star},(\hat{x}^{\prime})^{\star}]=[\hat{p}^{\prime},
\hat{x}^{\prime}]=-i \hbar \hat{1}^{\prime}\;,\end{equation}
which is precisely $(i\hbar\hat{1}^{\prime})^{\star}$.
All other commutators between $(\hat{1}^{\prime})^{\star}$,
$(\hat{x}^{\prime})^{\star}$ and $(\hat{p}^{\prime})^{\star}$ vanish
identically. Hence, the $\star$-operation constructed is compatible with
the structure of ${\cal A}^{\prime}_p$, and provides an involution on this
algebra.

Let us consider now other choices of representatives of
${\cal O}^{\prime}$.
Adopt, eg, the choice $\{\hat{1}, \hat{x}+\hat{t}(\hat{H})^2,\hat{p}\}$.
It follows from Eqs. (2.7,8) that
\begin{equation} \left(\hat{x}+\hat{t}(\hat{H})^2\right)^{\star}=
\left(\hat{x}+\hat{t}(\hat{H})^2\right)-2i\hbar\hat{H},\;\;\;\;\;\;
\hat{p}^{\star}=\hat{p},\;\;\;\;\;\;\hat{1}^{\star}=\hat{1}.\end{equation}
These reality conditions are of the type (3.2), with
$\hat{X}_a^{\,i}\hat{C}_i=\!-2i\hbar\hat{H}$ for
$\hat{U}_a=\hat{x}+\hat{t}(\hat{H})^2,$ vanishing otherwise.
In particular, assumption (3.6) is verified. Therefore, one can introduce
a $\star$-operation on ${\cal A}^{\prime}_p$ by applying Eq. (3.3) to the
present case. In this way, one recovers
the $\star$-relations (4.1,2), and thus the same involution on the algebra
of physical observables that was obtained above.

Choose now the operators $\hat{1}$, $\hat{x}+\hat{t}\hat{H}$ and
$\hat{p}$ as representatives of ${\cal O}^{\prime}$. The reality conditions
are then given by
\begin{equation} (\hat{x}+\hat{t}\hat{H})^{\star}=(\hat{x}+
\hat{t}\hat{H})-i\hbar\hat{1},\;\;\;\;\;\;\hat{p}^{\star}=
\hat{p},\;\;\;\;\;\;\hat{1}^{\star}=\hat{1}.\end{equation}
These reality conditions are of the form (3.7), and induce on
${\cal A}^{\prime}_p$ the $\star$-operation defined through Eq. (4.2) and
\begin{equation} (\hat{x}^{\prime})^{\star}=(\hat{x}^{\prime})
-i\hbar\hat{1}^{\prime}\;.\end{equation}
Since Eqs. (4.2) and (4.6) imply again relation (4.3), and
$(\hat{1}^{\prime})^{\star}$ commutes with $(\hat{x}^{\prime})^{\star}$ and
$(\hat{p}^{\prime})^{\star}$, the introduced $\star$-operation is
compatible with the commutators of the physical observables, and is therefore
an involution on ${\cal A}_p^{\prime}$. However,
this involution differs from that obtained in Eqs. (4.1,2). This
proves that the involution induced on ${\cal A}_p^{\prime}$ from
reality conditions depends on the particular selection of representatives
made for the complete set of physical observables under consideration.

Suppose that we can represent the $\star$-relations on
${\cal A}_p^{\prime}$ as adjoint relations on a Hilbert space of physical
states, as suggested by Ashtekar. From the involution provided by Eqs.
(4.1,2), we would then arrive at a quantum theory in which the
observable $\hat{x}^{\prime}$ would be self-adjoint.
On the other hand, the involution defined through Eqs. (4.2)
and (4.6) would lead to a quantum theory in which $\hat{x}^{\prime}$
would not be represented by a self-adjoint operator, so that it should
not correspond to a real physical observable of the system. This ambiguity
in the quantization can be nonetheless removed by insisting, for instance, on
that the real classical variable $x$ should be represented by the quantum
observable $\hat{x}^{\prime}$. One would thus expect that the spectrum of
$\hat{x}^{\prime}$ should be real to guarantee that this observable has
always real expectation values. Hence, $\hat{x}^{\prime}$ should be
self-adjoint. By itself, this condition supports the use of involution
(4.1,2) in the quantization, and elliminates other possible $\star$-relations
on ${\cal A}_p^{\prime}$, like, eg, relation (4.6).

To close this section, we will present an example in which the involution
induced on ${\cal B}^{\prime}$ via reality conditions is not compatible
with the structure of the algebra of physical observables. Let us consider
a physical system with a first-class constraint of the form $H=0$, where
$H\in$$I\!\!\!\,R$ is the momentum canonically conjugate to a certain variable
$t\in$$I\!\!\!\,R$. We will assume that the reduced phase space of the system
is the cotangent bundle over the unit circle $S^1$. As elementary variables,
we can choose the complex vector space spanned by $\{1, t, H,
c_{\theta}\equiv\cos{\theta},s_{\theta}\equiv\sin{\theta},
p_{\theta}\}$. Here, $\theta\in S^1$, and
$p_{\theta}\in$$I\!\!\!\,R$ is the momentum conjugate to $\theta$.
The reality conditions on the corresponding algebra ${\cal A}^{(\star)}$
of quantum operators are given by Eq. (2.7) and
\begin{equation} \hat{c}_{\theta}^{\star}=\hat{c}_{\theta},\;\;\;
\;\;\;\hat{s}_{\theta}^{\star}=\hat{s}_{\theta},\;\;\;\;\;\;
\hat{p}_{\theta}^{\star}=\hat{p}_{\theta},\;\;\;\;\;\;\hat{1}^{\star}=\hat{1}
\;.\end{equation}
Besides, since $\cos^2{\theta}+\sin^2{\theta}=1$, we will impose the
algebraic relation
\begin{equation} (\hat{c}_{\theta})^2+(\hat{s}_{\theta})^2=
\hat{1}\;.\end{equation}
A complete set of physical observables is  ${\cal O}^{\prime}\equiv
\{\hat{1}^{\prime},\hat{c}_{\theta}^{\prime},
\hat{s}_{\theta}^{\prime},\hat{p}_{\theta}^{\prime}\}$, the
prime denoting equivalence classes in ${\cal A}_p^{\prime}$. The only
non-vanishing commutators in ${\cal O}^{\prime}$ are
\begin{equation} [\hat{c}_{\theta}^{\prime},\hat{p}_{\theta}^{\prime}]=
-i\hbar\hat{s}_{\theta}^{\prime},\;\;\;\;\;\;\;
[\hat{s}_{\theta}^{\prime},\hat{p}_{\theta}^{\prime}]=
i\hbar\hat{c}_{\theta}^{\prime}.\end{equation}
In addition, relation (4.8) implies that the physical observables in
${\cal O}^{\prime}$ must satisfy
\begin{equation} (\hat{c}_{\theta}^{\prime})^2+(\hat{s}_{\theta}
^{\prime})^2=\hat{1}^{\prime}\;.\end{equation}

If one chooses $\hat{1}$, $\hat{c}_{\theta}$,
$\hat{s}_{\theta}$ and $\hat{p}_{\theta}$ as the
representatives of ${\cal O}^{\prime}$, the procedure explained in Section
III allows one to obtain an $\star$-operation on ${\cal B}^{\prime}$
(the free associative algebra generated by ${\cal O}^{\prime}$) which is
compatible with the commutators (4.9) and the algebraic relation (4.10),
and hence provides an involution on ${\cal A}_p^{\prime}$.
Let us select instead the representatives
${\cal O}\equiv \{ \hat{1},(\hat{c}_{\theta}
+\hat{t}\hat{H}),\hat{s}_{\theta}, \hat{p}_{\theta}\}$.
{}From Eqs. (2.7) and (4.7) (and the commutator of $\hat{t}$ and $\hat{H}$),
we get
\begin{equation}  \hat{1}^{\star}=\hat{1},\;\;\;\;\;\;(\hat{c}_{\theta}
+\hat{t}\hat{H})^{\star}=(\hat{c}_{\theta}+\hat{t}\hat{H})-i\hbar\hat{1},
\;\;\;\;\;\;\hat{s}_{\theta}^{\star}=\hat{s}_{\theta},\;\;\;\;\;\;
\hat{p}_{\theta}^{\star}=\hat{p}_{\theta}.\end{equation}
These reality conditions are of the type (3.7). Thus, we can apply the
results of Section III to arrive at an involution on ${\cal B}^{\prime}$
which is defined through the $\star$-relations (4.11), but imposed on
equivalence classes in ${\cal O}^{\prime}$.
However, such a $\star$-operation is incompatible with the algebraic
relation (4.10), because
\begin{equation} \left((\hat{c}_{\theta}^{\prime})^2+
(\hat{s}_{\theta}^{\prime})^2-\hat{1}^{\prime}\right)^{\star}=
\left(\hat{c}_{\theta}^{\prime}-i\hbar\hat{1}^{\prime}\right)^2
+(\hat{s}_{\theta}^{\prime})^2-\hat{1}^{\prime}\neq 0\;.\end{equation}
So, the involution introduced on ${\cal B}^{\prime}$ does not supply a
well-defined
involution on the algebra ${\cal A}^{\prime}_p$ of physical observables.
This example shows that the freedom in choosing repres\-entatives
of the complete set of physical observables is in general
restricted by the consistency of the algebraic structures with
the $\star$-operation constructed on ${\cal A}_p^{\prime}$.

\section{Conclusions and Further Comments}
\setcounter{equation}{0}

We have shown that, in systems with first-class constraints, the involution
defined on the algebra ${\cal A}^{(\star)}$ of quantum operators does never
project unambiguously to the algebra ${\cal A}_p^{\prime}$
of physical observables.
The reason for this is that the $\star$-conjugates
of all the representatives of any class of observables never belong
to the same equivalence class in ${\cal A}_p^{\prime}$.

We have also proved that, under sufficiently general circumstances,
it is never\-theless possible to obtain a well-defined involution on
${\cal A}^{\prime}_p$ via reality conditions by making a particular
choice of representatives for the equivalence classes of physical observables.
The procedure to arrive at this involution is the following.
One must first find a complete set of physical observables
$\{\hat{U}^{\prime}_a\}$ in ${\cal A}^{\prime}_p$, and select representatives
$\{\hat{U}_a\}$ of them such that their $\star$-conjugates
$\{\hat{U}^{\star}_a\}$ satisfy requirements (3.2) and (3.6), namely,
such that every $\hat{U}^{\star}_a$ belongs to the free associative algebra
generated by $\{\hat{U}_a\}$ up to an operator which, as well as its
$\star$-conjugate, vanish modulo quantum constraints. One can then
introduce an involution $\star$ in the free associative
algebra ${\cal B}^{\prime}$
by defining $(\hat{U}^{\prime}_a)^{\star}$ as the equivalence class
of the observable $\hat{U}^{\star}_a$ [see Eqs. (3.2,3)].
This involution on ${\cal B}^{\prime}$ straightforwardly supplies
an involution on ${\cal A}^{\prime}_p$, provided that the
constructed $\star$-operation is compatible with the commutation and
algebraic relations which exist between the physical observables in
the complete set $\{\hat{U}^{\prime}_a\}$.

The involution obtained in this way on ${\cal A}^{\prime}_p$ depends on
the selection of a complete set of physical observables and of
specific representatives for them. While these choices are severely
restricted by the consistency conditions explained above, there is in
general some freedom left, so that, by adopting different choices, one
may arrive at non-equivalent involutions on the algebra of physical
observables. This introduces an ambiguity in the quantization method
suggested by Ashtekar which has to be added to that existing in other steps
of the programme.$^2$ However, such an extra ambiguity, rather than being
a supplementary complication, may become an additional help when attempting
to complete the quantization. This is due to the fact that, given an
involution $\star$ on the algebra ${\cal A}^{\prime}_p$ and a certain
representation for ${\cal A}^{\prime}_p$ on a vector space $V_p$ of quantum
states, there is a priori no guarantee that there exists an inner product
on $V_p$ with respect to which the $\star$-relations on physical observables
are realized as Hermitian adjoint relations in the resulting Hilbert space.
Thus, if such an inner product does not exist for a particular involution
on ${\cal A}^{\prime}_p$, one can always try to induce a different
involution on this algebra via reality conditions, and see whether it is
possible to find then an inner product with the desired properties.

We notice, on the other hand, that the introduction of an involution on
${\cal A}^{\prime}_p$ amounts essentially to determine the $\star$-conjugate
to a complete set of physical observables. When one expects that a set of
this kind, or at least some of its elements, correspond classically to
real observables of the system, it is reasonable to assume that they
should be represented by self-adjoint operators. The involution defined
on ${\cal A}^{\prime}_p$ should therefore ensure that these operators
coincide with their $\star$-conjugates. These requirements clearly restrict
the admissible involutions on physical observ\-ables. Moreover, in the case
that this type of physical arguments would apply to a complete set in
${\cal A}^{\prime}_p$, one would fully specify the involution on this
algebra. In this way, one can use physical intuition to remove (either
partially or totally)\linebreak
the ambiguity encountered when inducing an involution
on the algebra of physical observables from reality conditions.

Finally, an alternative strategy to rule out such an ambiguity could consist
in adopting a specific procedure to induce the involution $\star$ on
${\cal A}^{\prime}_p$. A procedure of this type
might be, eg, the following.$^{11}$ Let us denote by
${\cal A}_s\subset {\cal A}_p$ the subalgebra formed by all
the strong quantum observables of the theory (that is, the operators which
commute strongly with all the quantum constraints $\{\hat{C}_i\}$), and
define ${\cal I}_s\equiv {\cal I}_C\bigcap {\cal A}_s$. It is immediate to
check that ${\cal I}_s$ is an ideal of ${\cal A}_s$. Suppose then that, in
the system under consideration, the involution $\star$ defined on
${\cal A}^{(\star)}$ and the representation constructed for the algebra
${\cal A}$ and for the constraints $\{C_i\}$ are such that:

a) The complex vector space spanned by the quantum constraints
$\{\hat{C}_i\}$ is closed\\ \indent under reality conditions, ie,
$\hat{C}_i^{\star}=\lambda_i^{\,j}\hat{C}_j$, where the
$\lambda_i^{\,j}$'s are complex numbers.

b) The algebra ${\cal A}_s^{\prime}\equiv {\cal A}_s/{\cal I}_s$
is isomorphic to ${\cal A}_p^{\prime}$.

c) The ideal ${\cal I}_s$ is invariant under the $\star$-operation.

\noindent Notice that hypothesis c) is in principle compatible with the
fact that ${\cal I}_C$ is not a $\star$-ideal of ${\cal A}_p$. Requirement
b), on the other hand, guarantees that each physical observable in
${\cal A}_p^{\prime}$ possesses (at least) one representative which is a
strong observable.

Using condition a), it is possible to prove that the $\star$-operation
leaves ${\cal A}_s$ invariant. Assumption c) ensures then that the
$\star$-relations project unambiguously to ${\cal A}^{\prime}_s$. One
hence obtains a well-defined involution on ${\cal A}_s^{\prime}$ which,
given condition b), supplies a unique involution on ${\cal A}_p^{\prime}$
through the existing isomorphism between these two algebras. So,
provided that hypotheses a)-c) are satisfied, the above strategy actually
allows one to induce an unambiguous involution on ${\cal A}_p^{\prime}$
from reality conditions.
\vspace*{.9cm}

{\bf Acknowledgments}

The author is greatly thankful to A. Ashtekar for helpful discussions and
valuable suggestions. He also wants to thank P. F. Gonz\'alez D\'{\i}az and
J. Mour$\tilde{\rm a}$o for useful
conversations. This work was supported by funds provided by DGICYT
and the Spanish Ministry of Education and Science under Contract
Adjunct to the Project No. PB91-0052.

\end{document}